\documentclass[journal,]{IEEEtran}

\pdfoutput=1
\usepackage[T1]{fontenc}
\usepackage{amsmath}
\usepackage[cmintegrals]{newtxmath}

\usepackage{colortbl}  
\usepackage{color}
\usepackage{array}   
\usepackage{graphicx} 
\usepackage{subfigure}
\usepackage{float} 

\usepackage{fancyhdr}

\graphicspath{{figure/}}

\begin{document}

\title{Integrating Communication, Sensing and Computing in Satellite Internet of Things: Challenges and Opportunities}

\author{Yong Zuo, Mingyang Yue, Huiyuan Yang, Liantao Wu, and Xiaojun Yuan

\thanks{
     Yong Zuo is with the National University of Defense Technology;
	 Mingyang Yue, Huiyuan Yang and Xiaojun Yuan are with the University of Electronic Science and Technology of China;  Liantao Wu is with the East China Normal University.
	 }
}

\maketitle

\IEEEpubid{\begin{minipage}{\textwidth}\ \\[50pt] \centering
\copyright 2023 IEEE. Personal use of this material is permitted. Permission 
from IEEE must be obtained for all other uses, in any current or future 
media, including reprinting/republishing this material for advertising or 
promotional purposes, creating new collective works, for resale or 
redistribution to servers or lists, or reuse of any copyrighted 
component of this work in other works.
\end{minipage}}

\IEEEpeerreviewmaketitle

\begin{abstract}
Satellite Internet of Things (IoT) is to use satellites as the access points for IoT devices to achieve the global coverage of future IoT systems, and is expected to support burgeoning IoT applications, including communication, sensing, and computing.  However, the complex and dynamic satellite environments and limited network resources raise new
challenges in the design of satellite IoT systems. In this article, we focus on the joint design of communication, sensing, and computing to improve the performance of satellite IoT, which is quite different from the case of terrestrial IoT systems.  We describe how the integration of the three functions can enhance system capabilities, and summarize the state-of-the-art solutions. Furthermore, we discuss the main challenges of integrating communication, sensing, and computing in satellite IoT to be solved with pressing interest. 
\end{abstract}

\section{Introduction}
The Internet of Things (IoT) enables a lot of devices to interact with environments or servers without human intervention, and is envisioned as an evolution of the Internet to connect everything. The IoT has enormous potential in various applications, such as smart city, environment monitoring, vehicle-to-vehicle communication, and unmanned aerial vehicles (UAVs), where the devices are 
dedicated to different tasks including communication, sensing, and computing. The connectivity of devices in the IoT  has been studied extensively, and  many wireless technologies have been developed, such as Sigfox, LoRa, and narrowband IoT (NB-IoT) \cite{centenaro2021survey}. In the fifth generation (5G) wireless communications, the 3GPP has defined a key application scenario named massive machine type communications (mMTC), which refers to the communication between a vast number of devices and a base station (BS). In mMTC, the access point (AP), e.g., the BS in a cellular system, is needed to manage the connections of the devices. However, due to commercial and engineering difficulties, it is impractical to deploy APs over remote areas, such as deserts, forests, and oceans. As such,  the coverage of terrestrial IoT is limited.

The low earth orbit (LEO) satellite is a potential enabler to achieve the global coverage of IoT. It is reported that the LEO satellite constellation is expected to contain thousands of satellites to provide global communication services. For example, the Starlink constellation is designed to consist of about 42,000 satellites in total, and more than 3,400 satellites have been launched. Compared with the existing terrestrial IoT systems, the satellite IoT naturally supports the coverage of remote areas, and is more robust in different terrestrial environments, such as floods and earthquakes where terrestrial APs fail to work. 
Besides the communication needs,  various sensing services, such as environment monitoring, geographical survey, and situational awareness, are expected to be realized by the satellite IoT. With high-speed motion and wide-band transmission, the satellite system is able to support remote sensing in real time. The widely distributed terrestrial sensors can also sense environmental objectives.  
In addition, the fast-growing artificial intelligence (AI), represented by deep learning, reinforcement learning, distributed learning, etc., has shown great potential for innovation in  various fields such as communication network optimization and intelligent sensing. To establish an intelligent system,  computing capability  is necessary for the satellite IoT system to deploy  AI. 
The satellite  equipped with high-performance central processing units (CPUs) can be seen as a server of mobile edge computing (MEC), to which the terrestrial devices can offload their computing tasks so as to enable AI and other computation-intensive and latency-critical applications.

\begin{figure}[t]
    \centering
    \includegraphics[width=3in]{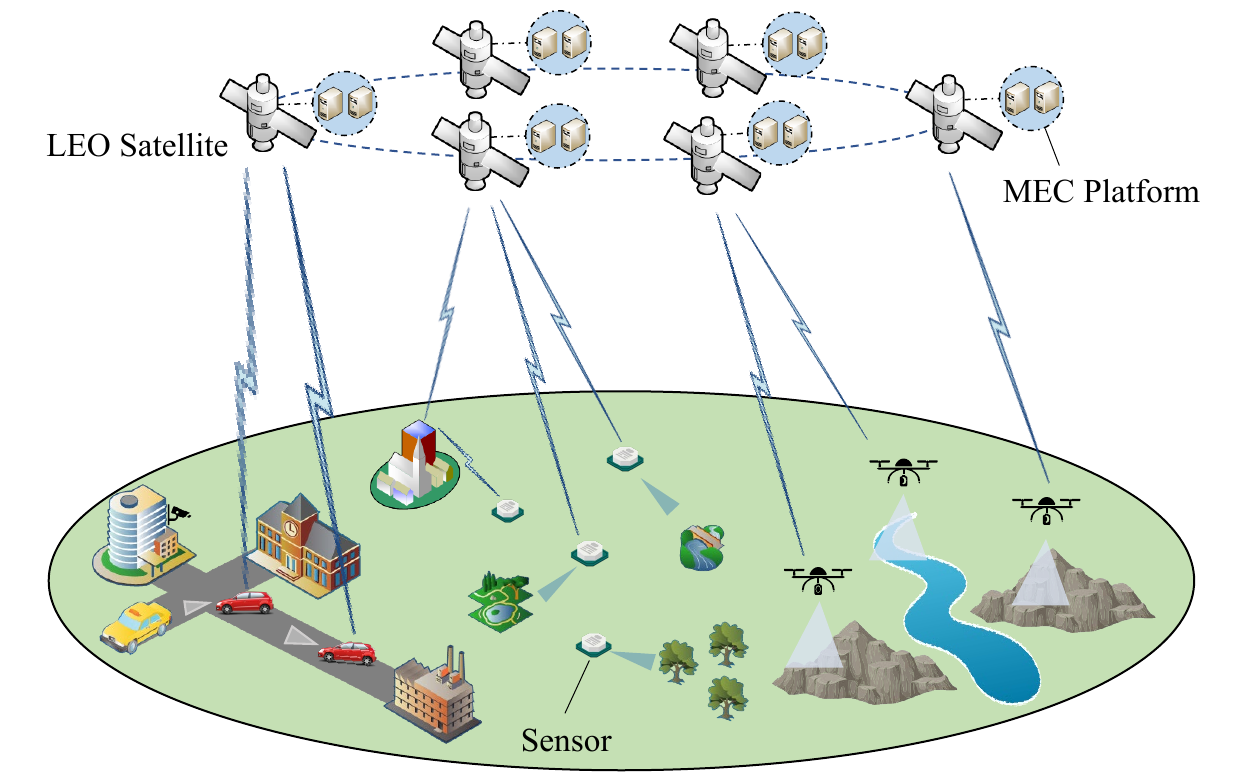}
    \caption{Scenario of integrating sensing, communication, and computing in  satellite IoT.}
    \label{system}
\end{figure}

The integration of communication, sensing, and computing has been widely considered as a future direction of terrestrial wireless networks \cite{6G}. Yet the integration of the three in satellite wireless networks has been much less studied.
An illustration of integrated communication, sensing, and computing in satellite IoT  is given in Fig. \ref{system}. With elaborate designs, the sensing and computing demand  can be better fulfilled with the help of communication. 
For example, a passive radar, which detects target objects by signals generated from other equipment, is able to utilize satellite communication signals to extend the detection range. 
As another example, we consider computation-intensive applications, such as autonomous vehicles. In these applications, the joint resource allocation of communication and computing can strike a favorable  balance between  latency  and energy consumption, compared with complete local processing  or complete offloading to the edge computing server.
As a third example, the intelligent sensing of UAVs can be supported by integrating communication and computing. In particular, based on the reinforcement learning framework,  UAVs interact with the environment, and transmit sensing data to  satellites with minimal communication and computing latency. The satellites  run reinforcement learning  and output a reward, based on which  UAVs  adjust the trajectory or sensing strategy to efficiently complete the sensing tasks. As such, it is highly desirable to integrate communication, sensing, and computing in satellite IoT networks.

The design of satellite IoT, however,  is very much different from  that of terrestrial IoT systems. Due to the large coverage, a satellite serves a much larger number of devices than a terrestrial AP does. Meanwhile, the LEO satellite constellation is dense and the coverage areas of different satellites may be overlapped, which necessitates satellite scheduling. This also implies that resource allocation and management in satellite IoT are more complicated than that in terrestrial IoT. In addition, due to the satellite's motion in orbit, the  communication environments are rapidly changing, e.g., from city to ocean. 
With the above features, the integration of communication, sensing, and computing for the satellite IoT faces many new challenges. In this article, we aim to describe  the state-of-the-art solutions for the integration of communication, sensing, and computing for satellite IoT, and explore the challenges that need to be solved in the future.

The remainder of this article is organized as follows. We discuss integrated sensing and communication for satellite IoT in the following section. Integrated communication and computing are then discussed. Subsequently, the task-oriented integration of sensing, communication, and computing in satellite IoT is discussed. 
The article is concluded in the final section.

\section{Integrated Sensing and Communication for Satellite IoT}
In this section, we mainly discuss how communication can enhance the sensing capability of a satellite IoT system. First, the objective of sensing is to acquire information from the physical environment, such as  air, clouds, and  moving objects. Second, the network state also needs to be sensed, such as sensing the distribution of tasks so as to allocate resources efficiently. In the following, we discuss the two sensing scenarios one by one.

\subsection{Environment State Sensing}
The existing literature mainly explores detection using the signals of communication satellites, as in the case of passive radar. In \cite{passive-ship}, the authors investigated the passive radar for ship detection using communication satellite signals, to improve the accessibility, coverage, and resolution of terrestrial systems. The terrestrial receiver acquires and processes the satellite signals reflected by the maritime targets, to estimate the range and velocity of targets. The authors presented a high-range-resolution method by utilizing the signals from different communication channels.  Ref. \cite{passive-starlink} considered using  Starlink satellites as the transmitters of passive radar systems to obtain a  high-range resolution. It is shown that the Starlink-based passive radar provides a maximum detection range of two orders of magnitude longer than that of global navigation satellite systems (GNSS) -based passive radar. 
However, to implement the environmental state sensing,  the following new challenges need to be solved.

{\bf{Waveform Design}}:
In the above-mentioned works, the realization of sensing is based on the existing communication signals, and the sensing performance is limited since the waveform is designed for communication. To improve the overall performance of both sensing and communication, the waveform needs to be redesigned under  the constraints of integrated communication and sensing, which provides additional flexibility and freedom for performance improvement. The trade-off between communication and sensing makes waveform design a highly challenging task.

{\bf{Beam  Management}}:
 Since the coverage areas of  satellites generally overlap,  terrestrial devices can receive signals from different satellites in communicating and sensing. Thus, the cooperation of beam management between satellites is necessary for the integration of communication and sensing.

\subsection{Network State Sensing}
Real-time monitoring of the network state is of critical importance to ensure the normal operation of the satellite IoT system. Generally speaking,  the network state includes the activity of the devices,  the network congestion, and the interference level, which are highly dynamic as the satellites orbiting the earth. For example, in satellite IoT systems, a large number of IoT devices are distributed over a wild area, and only a small fraction of them are active, i.e., need to communicate with the satellite. Thus, the activity of devices indicates the areas that need more communication and computation resources. Here we focus on   how to sense the activity of the devices in a satellite IoT system.

In a typical communication system, the activity of the devices can be acquired through random access protocols.  Each active device selects a preamble in a preamble pool and transmits it to the AP, and then the AP sends a response as a grant for subsequent data transmission and assigns physical resources to the device. In the satellite IoT, this conventional protocol  becomes very inefficient, since there is a large number of  devices and thus the latency caused by preamble collisions is severe. Grant-free non-orthogonal  multiple access (GF-NOMA) has been considered  a promising solution for massive machine-type connectivity, where active devices directly transmit pilots and data to the satellite without waiting for permission. As such, in each transmission frame, the AP needs to conduct device activity detection (DAD), channel estimation (CE), and data detection (DD), based on the received pilots. In \cite{GF-NOMA}, the authors proposed a Bernoulli-Rician message-passing algorithm with expectation maximization to address the joint DAD and CE problem for the GF-NOMA in LEO satellite IoT. The authors assumed that aided by terrestrial  base stations,  the Doppler frequency shifts of the satellite channels are completely compensated. 
This assumption  does not hold  in  remote areas where the terrestrial BS is not available, and thus  the Doppler shifts cannot be perfectly precompensated. 

 In a LEO satellite communication system, the maximum Doppler shift is typically  over tens of kilohertz. As such, the Doppler effect is a big challenge due to the high speed of LEO satellites relative to the ground.  How to guarantee accurate  activity detection of a huge number of devices in such a hostile communication environment is a challenging problem.

\begin{figure}
    \centering
    \subfigure{
    \begin{minipage}[t]{1\linewidth}
    \centering
    \includegraphics[width=2in]{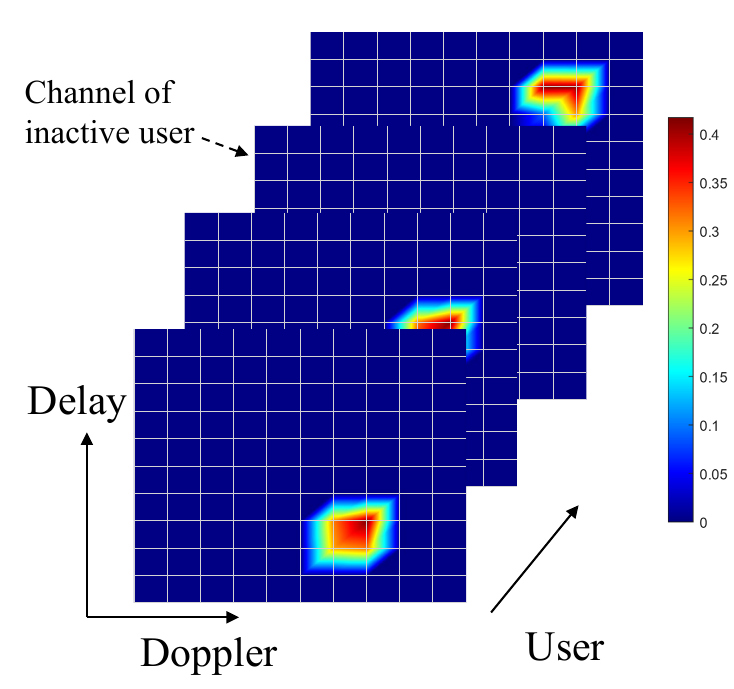}
    \end{minipage}
    \label{spar1}
        }    
    \subfigure{
    \begin{minipage}[t]{0.4\linewidth}
    \centering
    \includegraphics[width=1.5in]{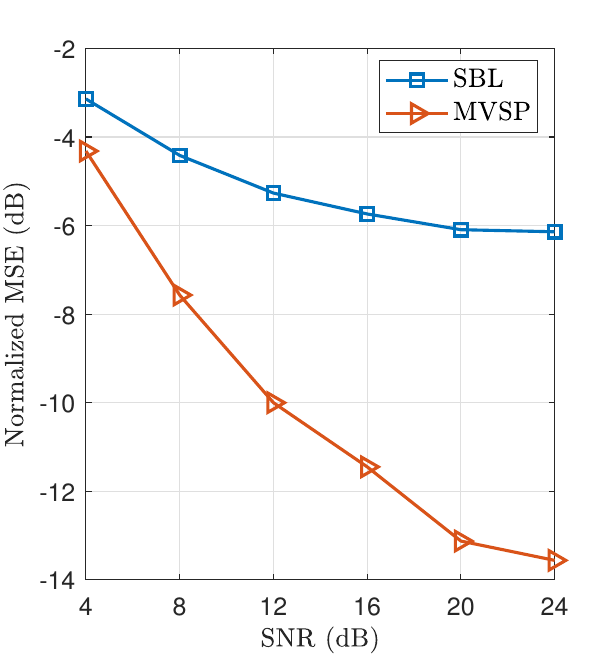}
    \end{minipage}
    \label{spar2}
    }\subfigure{
    \begin{minipage}[t]{0.4\linewidth}
    \centering
    \includegraphics[width=1.5in]{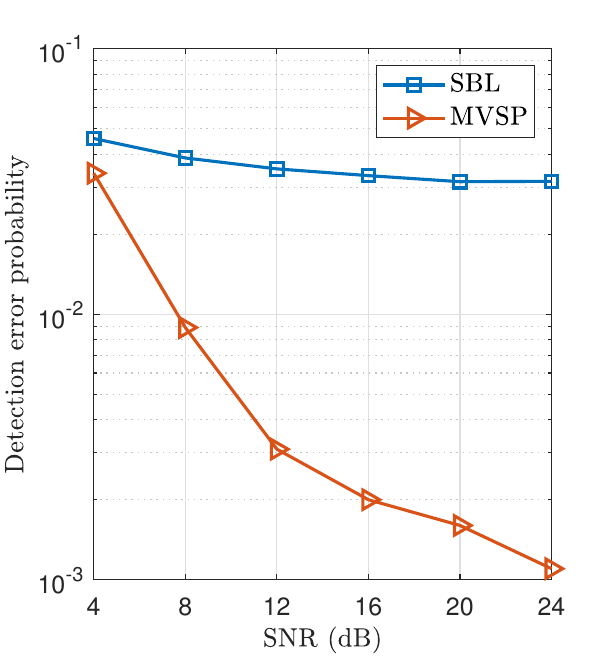}
    \end{minipage}
    \label{spar3}
    }
    \caption{An illustration of the channel sparsity structure, the channel estimation normalized mean square error (MSE) versus SNR, and the detection error probability versus SNR of a LEO satellite IoT system. We consider the SBL algorithm and the MVSP algorithm, where the latter exploits the three-dimension sparsity of the channel \cite{MVSP}. The average Doppler shift in a beam coverage is estimated and compensated. The left Doppler shifts vary from $[-15, 15]$ kHz with 2 kHz spreading.}
    \label{sparsity} 
\end{figure}

Exploiting the channel structure  is a possible solution. As shown in Fig. \ref{sparsity}, the channels of the satellite IoT scenario exhibit a sparsity structure in the delay-Doppler-user domain. The sporadic transmission pattern, where only a small fraction of the devices are active at any given time, results in sparsity in the user domain. The sparsity in the delay-Doppler domain is because the number of dominant paths of the satellite communication channel is limited, due to the weak multi-path transmission environment. Moreover, we see the block sparsity in the delay-Doppler domain, since the scattering effect causes delay and Doppler spread. From Fig. \ref{sparsity}, we see that by exploiting the channel sparsity in the delay-Doppler-user domain,  the modified variance state propagation (MVSP) algorithm \cite{MVSP} can significantly improve the performances of channel estimation and device activity detection as compared with the sparse Bayesian learning (SBL) algorithm.

\section{Integration of Communication and Computing for Satellite IoT}
The integration of communication and computing is a fundamental problem to achieve the full potential of satellite IoT systems. Recent studies show that the performance optimization of user services in satellite IoT systems, such as computation-intensive and latency-critical applications, critically depends on joint computing and communication techniques; see, e.g., the minimization of the weighted sum energy consumption in \cite{compu-solu1} and the minimization of the execution latency of the computation-intensive applications in \cite{compu-solu2}.

\subsection{State-of-the-Art Solutions}
Benefiting from the broad coverage and flexible deployment, satellite IoT systems have gained lots of attention in several scenarios, such as remote sensing,  vehicle-to-vehicle communication in remote areas and so on. In these scenarios, with increasing demands for high quality of services (QoS), user terminals with limited computation and storage capacities will be required to provide high reliability and low latency services for computation-intensive applications. Thus, allowing user terminals to offload the computation-intensive and latency-critical applications to the powerful satellite cloud server is a feasible option.

\begin{figure}
    \centering
    \subfigure[]{
    \begin{minipage}[t]{1\linewidth}
    \centering
    \includegraphics[width=3.3in]{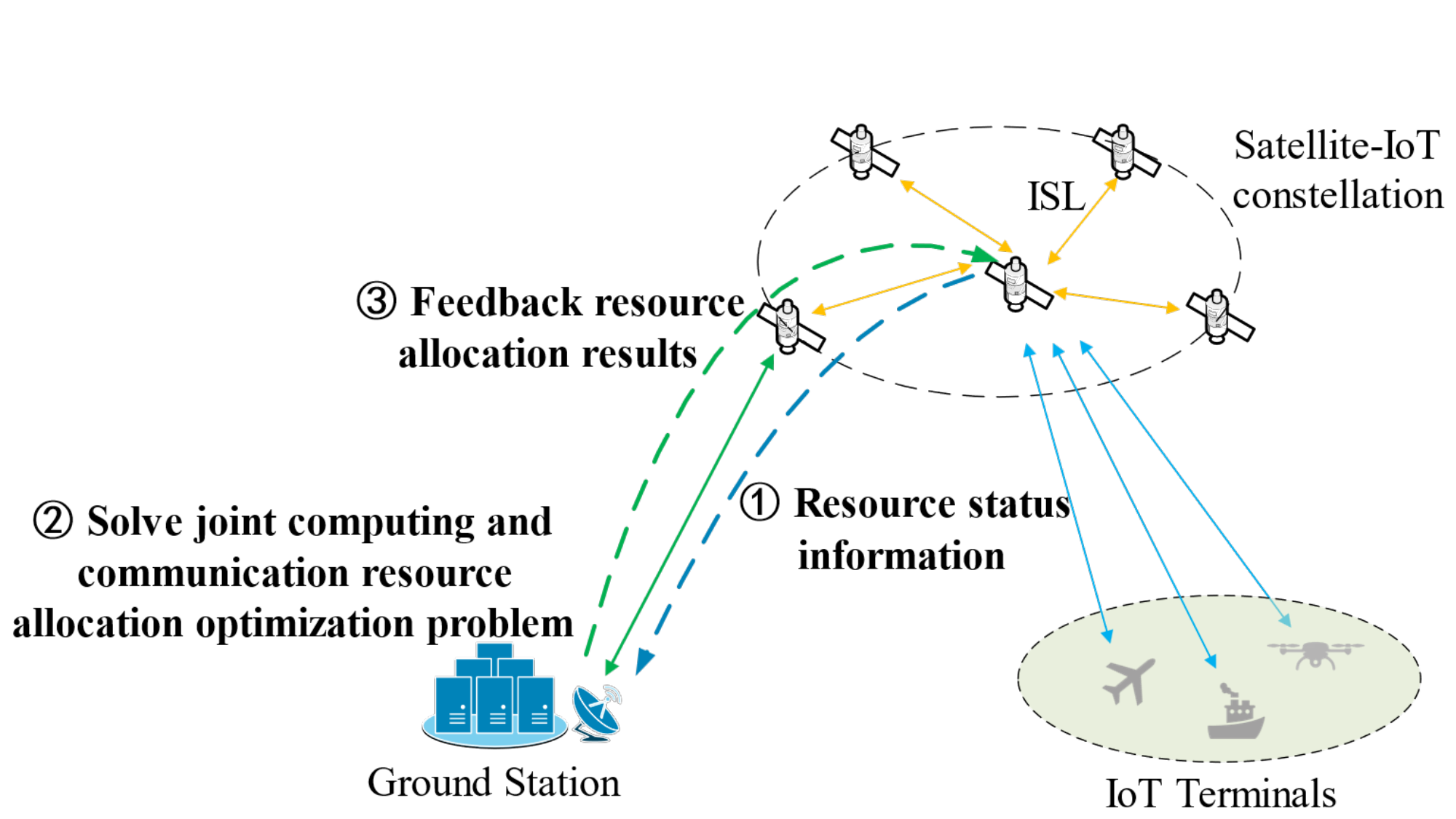}
    \end{minipage}
    \label{Sec3-1}
    }
    \subfigure[]{
    \begin{minipage}[t]{1\linewidth}
    \centering
    \includegraphics[width=3.3in]{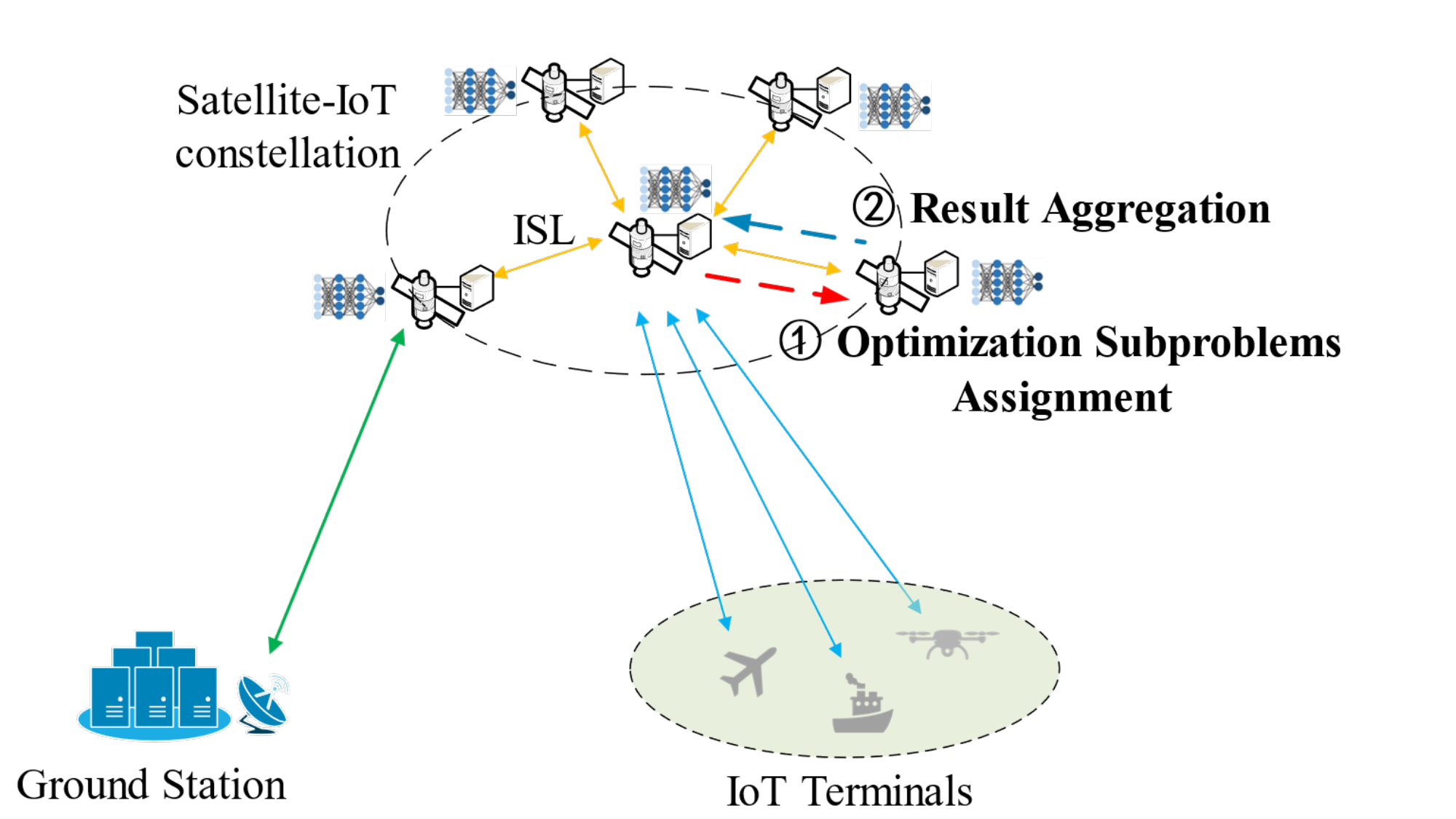}
    \end{minipage}
    \label{Sec3-2}
    }
    \caption{Solutions of integrated communication and computing for satellite IoT: a) Centralized system with better performance and higher link information transmission delay; b) Distributed system with performance loss and faster resource allocation decision time.}
    \label{Sec3Fig} 
\end{figure}

Fig. \ref{Sec3Fig} shows two solutions of integrated communication and computing for satellite IoT, i.e., a centralized system and a distributed system. In the centralized system, the ground station solves the joint computing and communication resource allocation problem based on the satellite resource status information, and then the allocation result is transmitted to the satellites. 
Instead, in a distributed system, the allocation problem is divided into several sub-problems which are assigned to different satellites but not ground stations. Without the large-link transmission delay between the satellite and the ground station, the distributed system has less resource allocation decision time than the centralized system. However,  the distributed system generally suffers from performance loss due to the conversion of the original problem into distributed sub-problems.

At present, the integration of communication and computing in satellite IoT systems is still in its infancy. The basic idea is to construct and solve a mathematical optimization problem about joint computing and communication resource allocation under some constraints, such as maximum tolerable delay, maximum transmit power, and total computation resource. Specifically, the overall tolerable delay of a service, which consists of the offloading time delay of the radio access network (RAN) and the onboard computation time delay, serves as a constraint  of joint communication and computing optimization in satellite IoT systems.
In \cite{compu-solu1}, a satellite-aerial integrated edge computing network combines a low-earth-orbit (LEO) satellite and aerial high-altitude platforms (HAPs) to provide edge computing services for ground user equipment (GUE). In the system, GUE’s computing tasks can be offloaded to HAP(s) or LEO satellites. The authors minimize the weighted sum energy consumption of satellite-aerial integrated edge computing networks (SAIECN) via joint GUE association, multi-user multiple-input  multiple-output (MU-MIMO) transmit precoding, computation task assignment, and resource allocation. To address this challenging problem,  the authors of \cite{compu-solu1} decomposed the optimization problem into four sub-problems and solve them  recursively.
In \cite{compu-solu2}, the authors investigated the joint computing and communication resource management problem for satellite IoT systems to minimize the execution latency of the computation-intensive applications, where two edge computing scenarios and a complete local execution scenario are considered. Furthermore, the joint computing and communication resource allocation problem for the computation-intensive services is formulated as a mixed-integer programming problem which is solved by a game-theoretic and many-to-one matching theory based scheme. Compared to \cite{compu-solu1}, the algorithm in \cite{compu-solu2} is more promising since its resource allocation problem is convex for a given vector of offloading decisions and association indicators and the Karush-Kuhn-Tucker (KKT) conditions can be adopted.

\begin{figure}[htbp]
    \centering
    \includegraphics[width=3.4in]{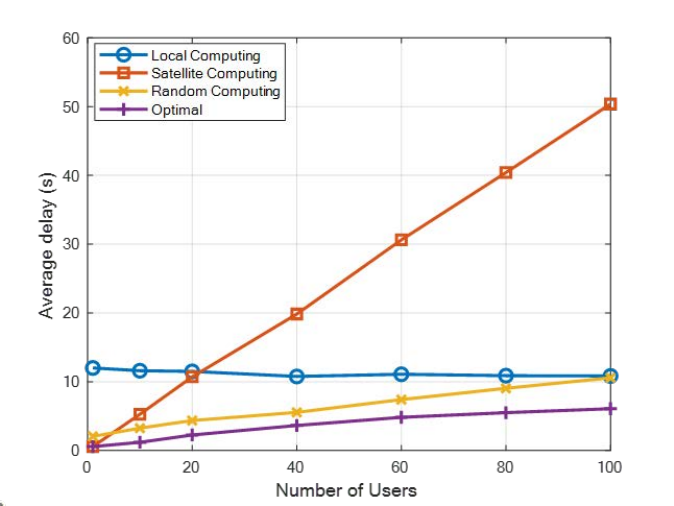}
    \caption{Task execution delay comparison versus different numbers of users using local computing, satellite computing, random computing and optimal task offloading with the cross entropy method. We consider 4 satellites with  a transmission rate of 1 Gbps and a computation capacity of 10 GHz, where the computation capacities of  devices are randomly selected from the set $[0.3, 0.4, 0.5, 0.6, 0.7]$ GHz. The task data size  follows the uniform distribution with $[500, 5000]$ KB. The processing density of
    the computing task is randomly chosen from $[500,2000,3000]$ cycles/bit.}
    \label{sec3sim}
\end{figure}

The task execution delay comparison versus different numbers of users using different methods is shown in Fig. \ref{sec3sim}. As the computing capability of a satellite is limited, when the number of users increases, allocated computing resources are decreased, leading to a larger task delay. Thus, when the number of users is small, satellite computing achieves a smaller delay than local computing.  An increasing number of users leads to fewer users choosing to offload tasks to satellite rather than to process tasks on a local device, and this results in the rise of the average delay. Local computing  does not rely on the resources of satellites, and thus it keeps unchanged.
The optimal task offloading with the cross entropy method, which has been shown to be efficient in finding near-optimal solutions to complex combinatorial optimization problems, achieves the smallest delay with a varying number of users.

\subsection{Research Challenges}
The optimal allocation  of communication and computing resources in satellite IoT systems is an emerging research direction rich in open challenges, as discussed below. 

\textbf{Service Continuity:} Due to the rapid movement of the LEO satellites, a user device may frequently switch its accessed satellite.  How to ensure service continuity is a key issue when a user device moves from one satellite MEC platform to another. Service migration is an effective solution. During the service migration, the original platform sends the application and data to the new platform. The key to service migration is to determine when and where to migrate. For migration timing,  the best practice is premigration, where the migration is completed as the user device enters the new service area. The migration target satellite MEC platform can be predicted since the topology changes of a satellite network are periodic. If the prediction of the target satellite MEC platform is accurate,  the migration can maintain service continuity and reduce the communication delay between the user device and the MEC platform. However, if the prediction is wrong,  the user experience will be significantly impacted and even the service may be interrupted.

\textbf{Uneven Service  Distribution:} Because a satellite covers a relatively small area in satellite IoT systems, traffic requirements are unbalanced due to the varying population density, which is high in cities, low in rural areas, and almost zero over the oceans (which cover about 70 percent of the surface of the earth). As satellites move, traffic received from terrestrial nodes varies continuously based on the user density in the footprint area, resulting in the unbalanced utilization of communication and computing resources between different satellites. Then, how to balance the uneven utilization of communication and computing resources to improve the network efficiency (e.g., in terms of task execution delay)  becomes an essential problem for satellite IoT systems \cite{compu-2}. The dynamic scheduling algorithm is expected to balance onboard resources between different satellites based on service requirements monitoring.

\textbf{Task Scheduling Diversity:} With the emergence of inter-satellite links, satellite clustering has become a new trend in the development of satellite networking. Satellites equipped with computing and storage resources form the satellite edge computing cluster, which can realize the migration of computing resources to the edge of the network \cite{compu-1}. With the support of inter-satellite links and mobile edge computing, satellite edge computing clusters will be more efficient and intelligent. However, the task scheduling decision factors of the satellite IoT systems are very diverse and flexible, which makes the task scheduling problem difficult to solve. The mobility of satellites constitutes a major difference between satellite networks and their terrestrial counterparts.  There are two types of decisions to be made in satellite edge computing:
\begin{enumerate}
    \item Using a single satellite or multiple satellites to serve one user depending on computing and communication resources, service distribution, QoS requirements, and so on.
    \item Cooperative computation offloading modes selection, including local processing mode, satellite edge computing mode, and cooperative computation mode between local user and satellite.
\end{enumerate}

\textbf{Distributed Algorithm Design:} For satellite IoT systems, it is usually quite difficult to solve the mathematical optimization problem about joint computing and communication resource allocation \cite{compu-solu1,compu-solu2}.  In addition, various design criteria in terms of maximum tolerable delay and maximum transmit power shall be considered in the problem formulations. In Fig. \ref{sec3sim}, solving the non-convex problem  by decomposing the optimization problem into some sub-problems and solving them iteratively involves high computational complexity and requires centralized computation in the ground station. A straightforward approach is to design a low-complexity and distributed algorithm executed on the satellite so as to strike a balance between the global loss and completion time.

\section{Task-Oriented Integration of Sensing, Communication, and Computing In Satellite IoT}

Many industries, such as transportation, farming, and environment monitoring, need to provide services by deploying sensing devices in remote areas. In consideration of computing capacity, energy consumption, and delay, these sensing devices generally need to transmit the sensing data to servers for processing. Since terrestrial networks are unavailable in these remote areas, adopting satellite-based technologies becomes an attractive alternative. We refer to this type of sensing scenarios as satellite-aided sensing. In the following two subsections, we demonstrate the advantages of integrating sensing, communication, and computing by taking two satellite-aided sensing scenarios as examples.

\subsection{Task-Oriented Communication in Satellite IoT}

\begin{figure}[]
	\centering
	\includegraphics[width=3in]{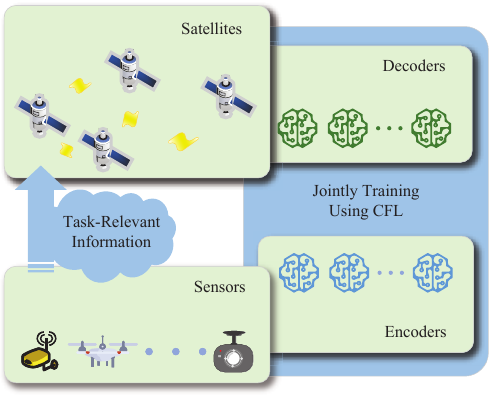}
	\caption{Task-oriented sensing data transmitting in satellite IoT.}
	\label{fig_semantic}
\end{figure}

Consider the scenario in Fig. \ref{system}, where multiple terrestrial sensing devices in remote areas independently perform a common sensing task (e.g., many UAVs independently perform surface mapping tasks) aided by several servers (deployed on satellites). To accomplish the task cooperatively, these sensing devices need to share data with the servers. A common practice for the sensing devices is to upload the raw sensing data to the satellites. 

For many sensing tasks, the sensing data to be transmitted are high-dimensional with a high generation rate. Directly transmitting raw data to the satellites may incur unbearably high communication costs and latencies. Universal compression methods, lossless or lossy, are often adopted to reduce the amount of data transmitted. However, since these compression methods are usually designed for general purposes, they tend to retain much task-irrelevant information but discard much task-relevant information\footnote{Task-relevant information refers to information useful for completing the task.}, leading to a mismatch with the specific task and thus causing a significant performance penalty.

To reduce communication cost and latency, we wish to transmit only the task-relevant information, even though some computing resources may be paid as a cost. Task-oriented communication \cite{shao2021learning} is a promissing technique available to fulfill this demand. Task-oriented communication is essentially a deep learning-based lossy compression technique: an encoder and a decoder (both neural networks) are jointly trained to enable the encoder to extract task-relevant information from raw data. In this way, only the relatively low-dimensional features containing the task-relevant information need to be transmitted.\footnote{For instance, for the objective classification tasks, the output of the encoder can feed into the decoder to form a complete neural network, which is then trained on a dataset of labeled objective images. After training, the encoder and decoder are respectively deployed to the sensing device and server. When server support is required, the sensing device feeds raw sensing data into the encoder, which outputs low-dimensional features containing task-relevant information. These features are transmitted to the server on satellites and then fed into the decoder to obtain the information required to complete the task.} Clearly, as a lossy compression technique, relative to lossless compression, task-oriented communication inevitably leads to a degradation of sensing task performance. However, preliminaries suggest that, for some sensing tasks, an acceptable performance penalty can be exchanged for a thousand-fold saving in communication resources \cite{shao2021task}. Furthermore, the performance penalty can be largely reduced by consuming more computing resources, e.g., by resorting to more powerful but typically computationally more demanding neural networks.
This indicates the trade-off between computing resource consumption and sensing task performance.

How to efficiently train the encoder and decoder in satellite IoT is also a critical challenge.  Recall that we consider the scenario where several sensing devices independently perform identical tasks. This implies that these sensing devices can, and should, collaboratively train the same pair of encoder and decoder. Since the computing resources are enriched at the servers on satellites, a straightforward approach is to upload the sensing data to the satellites for centralized training. Obviously, this introduces undesired communication costs. Federated learning (FL) \cite{Yang2022Federated} seems to be an attractive alternative. Indeed, FL realizes collaborative training of machine learning models through gradient/model exchange, thereby avoiding raw data uploading. 
Nevertheless, in satellite IoT, multiple satellites form a decentralized network, and each satellite is responsible for the access of a group of IoT devices.  It goes against the traditional FL paradigm, where a server is endowed with a central authority to coordinate all the participating devices.
We suggest applying confederated learning (CFL) \cite{wang2022confederated} to enable collaborative training in satellite IoT. CFL is a hierarchical version of FL, where the upper layer is a decentralized network formed by several CFL servers, and the lower layer is composed of some device sets, with the devices in each set connected to a single server. In satellite IoT, each satellite and its corresponding IoT devices can be a CFL server and its connected devices, respectively. During each training round, each satellite server first acquires local models from its connected IoT devices, aggregates them together with the models from other neighboring satellite servers in the constellation, and then broadcasts the aggregated model to its connected IoT devices. CFL fits well with our considered scenario since a satellite constellation is by design a decentralized network. On the other hand,  CFL retains the advantage of FL, where only models are required to be transmitted. Besides, we note that the sensing data is not required to be shared with the server during training and inference (i.e., perform the sensing tasks with task-oriented communication). This is a valuable feature for privacy-sensitive tasks. The envisaged task-oriented sensing data transmitting system is summarized in Fig. \ref{fig_semantic}.

\subsection{Task-Oriented Sensing in Satellite IoT}

In this subsection, we briefly introduce task-oriented sensing in satellite IoT, which is another possible application case of deep integration of sensing, communication, and computing in satellite IoT. We still consider the scenario in Fig. \ref{system}, except that the sensing devices are further required to learn how to sense. This is reasonable when, for instance, the sensing task is newly proposed or the sensed environment is newly encountered. For such a problem, a natural solution is to apply the reinforcement learning (RL) technique. Specifically, let the sensing device be the RL agent, the sensing strategy be the RL action, and the sensed environment be the RL state. The RL agent uses long-term return, a weighted sum of accumulated rewards, to evaluate the quality of the policy (a mapping from states to actions). In our scenario, we assume that the computing capacity of the sensing device is insufficient to calculate the reward in real time;
thus, the RL agent needs to share data with servers on satellites, and the latter then returns the corresponding task-specific rewards. Evidently, we can still use task-oriented communication introduced in the previous subsection to avoid transmitting raw sensing data to reduce communication costs. The above RL-based task-oriented sensing framework is an organic integration of sensing, communication, and computing. The integration with computing results in reduced communication costs and improved sensing capability in satellite IoT; meanwhile, the consumption of communication resources also improves sensing efficiency.

There are some difficulties when deploying RL in the satellite IoT.  For example,  the communication encoder and decoder need to be retrained every once in a while due to the following two reasons.  1) In the RL process, the RL policy constantly changes, causing the distribution of the sensing data changes accordingly; 2) the communication encoder and decoder work well only when they adapt to the data distribution.
For this problem, resorting to federated incremental learning \cite{Dong2022Federated} is a possible solution.

\section{Conclusion}
This article discussed how the integration of communication, sensing, and computing enables satellite IoT. We explained how the integration improves the system performance, discussed the state-of-the-art solutions, and pointed out research challenges.   Other IoT systems, such as the high altitude platform (HAP)  and  geostationary (GEO) satellite enabled IoT systems, can also be enhanced by the integration of communication, sensing, and computing. In conclusion, this article serves as a humble attempt to provide useful guidance and insightful inspiration for future research endeavors on satellite IoT systems.

\section{Acknowledgment}
The work of Yong Zuo was supported by the National Key Research and Development Program of China under Grant 2020YFB1804800. The work of Mingyang Yue, Huiyuan Yang, and Xiaojun Yuan was supported in part by the National Natural Science Foundation of China under Grant 62071090, and in part by the Sichuan Science and Technology Program under Grant 2022ZYD0120. The work of Liantao Wu was supported by the National Natural Science Foundation of China under Grant 62202307.
\bibliographystyle{IEEEtran}
\bibliography{reference}

\section{Biography}
Yong Zuo [M’22] (zuoyong@nudt.edu.cn) received his Ph.D. degree in communication engineering from the Chinese Academy of Sciences. He is now a professor with the National University of Defense Technology. He is also an Adjunct Professor with Xiangjiang Laboratory and Zhejiang Lab.

Mingyang Yue (myyue@std.uestc.edu.cn) received his B.S. degree in communication engineering from the University of Electronic Science and Technology of China in 2020, where he is currently pursuing a Ph.D. degree.

Huiyuan Yang [S’23] (hyyang@std.uestc.edu.cn) received his B.S. degree in communication engineering from the University of Electronic Science and Technology of China, Chengdu, China, in 2019, where he is currently pursuing a Ph.D. degree.

Liantao Wu (wult@shangahitech.edu.cn) received his Ph.D. degree from Zhejiang University, China, in 2017. He is currently an Associate Professor with East China Normal University, China.

Xiaojun Yuan [S’04, M’09, SM’15] (xjyuan@uestc.edu.cn) received his Ph.D. degree in electrical engineering from the City University of Hong Kong. He is now a professor with the University of Electronic Science and Technology of China.

\end{document}